\documentclass[runningheads]{llncs}
\usepackage{graphicx}
\usepackage{multirow}
\usepackage[normalem]{ulem}
\usepackage{amsmath}
\begin{document}
%
\title{Sentence Compression as Deletion with Contextual Embeddings}
%
%
\author{Minh-Tien Nguyen\inst{1,2} \and Bui Cong Minh\inst{1} \and Dung Tien Le\inst{1} \and Le Thai Linh\inst{1,3}}


%
\authorrunning{Nguyen et al., accepted by ICCCI 2020}

%

\institute{CINNAMON LAB, \\10th floor, Geleximco building, 36 Hoang Cau, Dong Da district, Hanoi, Vietnam
\email{\{ryan.nguyen, matthew, nathan, linhlt\}@cinnamon.is}
\and
Hung Yen University of Technology and Education, Vietnam\\
\email{tiennm@utehy.edu.vn} \\
\and The University of Queensland, Australia}

\maketitle              
\begin{abstract}
Sentence compression is the task of creating a shorter version of an input sentence while keeping important information. In this paper, we extend the task of compression by deletion with the use of contextual embeddings. Different from prior work usually using non-contextual embeddings (Glove or Word2Vec), we exploit contextual embeddings that enable our model capturing the context of inputs. More precisely, we utilize contextual embeddings stacked by bidirectional Long-short Term Memory and Conditional Random Fields for dealing with sequence labeling. Experimental results on a benchmark Google dataset show that by utilizing contextual embeddings, our model achieves a new state-of-the-art F-score compared to strong methods reported on the leader board.

\keywords{Sentence compression  \and Summarization \and Transformers.}
\end{abstract}
\section{Introduction}\vspace{-0.1cm}
Sentence compression is a standard task of natural language processing (NLP), in which a long original sentence is compressed into a shorter version. The compression tries to retain important information, which can be used to reflect the original sentence. Table \ref{tab:exp} shows an example of compression.\vspace{-0.3cm}

\begin{table}[!h]
\centering
\setlength{\tabcolsep}{4pt}
\caption{An example of sentence compression. We can observe that the compression is created by deleting unnecessary words from the original sentence.}\label{tab:exp}
\begin{tabular}{p{1.7cm} p{9.7cm}} \hline
Original sentence & Floyd Mayweather is open to fighting Amir Khan in the future, despite snubbing the Bolton-born boxer in favour of a May bout with Argentine Marcos Maidana, according to promoters Golden Boy. \\ \hline
Compression & Floyd Mayweather is open to fighting Amir Khan in the future. \\ \hline
\end{tabular}\vspace{-0.4cm}
\end{table}
The output of compression models can be used in NLP systems, e.g. compression of summarization \cite{Vanetik_IS_20} or dialog summary generation \cite{Liu_SIGKDD_19}.

Over two decades, there are a lot of studies focusing on sentence compression. Although compression may differ lexically and structurally from the source sentence, there are two main approaches to this task. The first approach is extractive by deleting unimportant tokens from the original sentence \cite{Clarke_JAIR_08,Kirkpatrick_ACL_11,Filippova_EMNLP_13,Filippova_EMNLP_15,Wang_ACL_17,Zhao_ACL_18,Vanetik_IS_20}. The tokens can be defined as words \cite{Filippova_EMNLP_15,Wang_ACL_17,Zhao_ACL_18,Vanetik_IS_20} or phrases from parsed trees \cite{Kirkpatrick_ACL_11,Filippova_EMNLP_13}. By contrast, the abstractive compression approach crates compression by paraphrasing tokens from the original sentence. Therefore, tokens of the compression do not need to be similar to those in the original sentence. To do that, many neural machine translation models can be utilized \cite{Cho_EMNLP_14}. However, the quality of abstractive compression models is still far from human satisfaction.

In this paper, we study the task of sentence compression by deletion. The idea of our study comes from the fact that compression can be created by removing unnecessary tokens \cite{Filippova_EMNLP_13,Filippova_EMNLP_15,Wang_ACL_17,Zhao_ACL_18}. To do that, we formulate the compression as a sequence labeling task and introduce a sequence labeling model. More precisely, inspired by the recent success of contextual embeddings \cite{Akbik_COLING_18,Devlin_NAACL_19}, we employ contextual embeddings as word embeddings. The contextual concept enables our model to exploit word contextual transformation for word representation. To learn hidden representation, we stack bidirectional Long-short Term Memory (BiLSTM) on the embedding layer. The sequence tagging is done by using Conditional Random Fields (CRFs) to take into account global optimization over the whole sequence. This paper makes two main contributions as the following:
\begin{itemize}
    \item It introduces a neural-based deletion model for sentence compression. The model exploits contextual embeddings from transformers, which allow our model to encode the contextual aspect of input words. It efficiently facilitates learning the hidden representation of data.
    
    \item It validates the efficiency of the model on the Google dataset, a benchmark corpus of sentence compression. Experimental results show that our model with contextual embeddings achieves a new state-of-the-art F-score compared to previous strong methods reported on the leader board.
\end{itemize}{}

We review related work of compression in Section \ref{sec:related_work}. We next introduce our model, including data preparation, layers, and the training process in Section \ref{sec:model}. Settings and evaluation metrics are showed in Section \ref{sec:setting}. Comparison and observation are reported in Section \ref{sec:result} . We finally draw conclusions in Section \ref{sec:conclusion}.\vspace{-0.3cm}

\section{Related Work}\label{sec:related_work}\vspace{-0.1cm}

The compression task has been addressed in two main directions: deletion and abstraction. The deletion approach usually treats the compression as an extraction or a sequence labeling task, in which unnecessary words or tokens are removed. The removal can be done on several level of word units \cite{Clarke_JAIR_08,Kirkpatrick_ACL_11,Filippova_EMNLP_13,Filippova_EMNLP_15}. For example, Kirkpatrick et al. introduced a model for jointly learning sentence extraction and compression for multi-document summarization \cite{Kirkpatrick_ACL_11}. The model used $n$-grams and compression features to score extraction candidates. The model infers to extract candidates by using Integer Linear Programming (ILP). By jointly learning, the model achieved the highest ROUGE-scores of TAC 2008. Filippova and Altun addressed the lack of data for sentence compression by introducing a new dataset \cite{Filippova_EMNLP_13}. The dataset was created by using deletion-based algorithms on hundreds of thousands of instances. To do that, the authors used ILP on syntactic trees. In extension work, Filippova et al. employed Long-short Term Memory (LSTM) for word deletion \cite{Filippova_EMNLP_15}. The model used LSTM to learn the hidden representation of input tokens. Labels of tokens in the final layer were predicted by using softmax. Experimental results show that this simple model is better than baselines in terms of readability and informativeness. Wang et al. improved the domain adaptability of a deletion-based LSTM for sentence compression \cite{Wang_ACL_17}. The authors assumed that syntactic information helps to make a robust model. To do that, the authors defined syntactic features and introduced syntactic constraints solved by ILP. The evaluation shows that this method is better than a traditional non-neural-network in a cross-domain setting. Zhao et al. exploited a neural language model as an evaluator for deletion-and-evaluation \cite{Zhao_ACL_18}. To do that, a reward function was defined by using a series of trial-and-error deletion operation on the source sentence to obtain the best target compression. Vanetik et al. presented an unsupervised constrained optimization method for summarization compression by iteratively removing redundant parts from original sentences \cite{Vanetik_IS_20}. The model used constituency-based parse trees and hand-crafted rules for creating elementary discourse units. To do that, the authors defined a weighted function computed by a parse tree gain for assigning weights into tokens. Experimental results confirm the efficiency of the model in the task of single-document summarization. The relation of our model to existing work is that we share the idea of sentence compression as a deletion task. However, we also enrich the task by introducing a model based on contextual embeddings.

There is little research on abstractive compression. A very close research direction is machine translation, in which a translator receives an original sentence and translates it into a shorter version. From this formulation, we can utilize many sophisticated neural machine translation (NMT) models for compression as a sequence-to-sequence problem \cite{Cho_EMNLP_14}. However, even with the recent success of abstractive compression models, their quality is still quite far from human expectation. Therefore, we focus on sentence compression by deletion.\vspace{-0.2cm}


\section{Sentence Compression with Contextual Embeddings}\label{sec:model}
\begin{figure}[!h]
    \centering
	\includegraphics[width=0.9\textwidth]{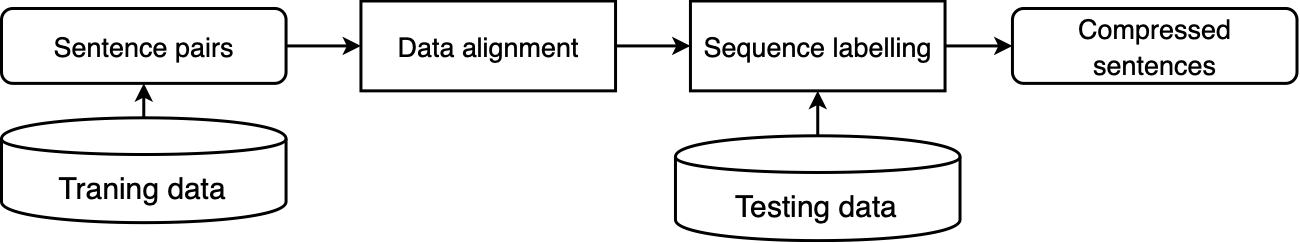}
	\caption{The process of our model.}\label{fig:general-model}\vspace{-0.4cm}
\end{figure}

\vspace{-0.1cm}
This section shows our proposed model of the sentence compression task. We introduce the general process of our model in Fig. \ref{fig:general-model}.\vspace{-0.3cm}
The data alignment prepares training data for training the sequence labelling model. After training, the trained model is applied to testing samples to compress raw sentences. We first describe data preparation and next introduce our compression model. 

\subsection{Data Preparation}
\subsubsection{Dataset}
We used the Google dataset\footnote{http://nlpprogress.com/english/summarization.html} for the sentence compression task \cite{Filippova_EMNLP_13}. To create pairs of sentences, the headline and the first sentence of each article collected from Google News service were extracted. Instead of directly using headline-first-sentence pairs, the authors used the headline to find our proper extractive compression of the sentence by a matching algorithm. This is because headlines are syntactically different from the first sentences. So directly using headlines makes a challenge for compression by deleting words. We used a new version of the dataset. It includes automatic generated 200,000 pairs for traing and 10,000 pairs generated by humans for testing. Each pair contain an original sentence and a shorter version. Table \ref{tab:data} shows data observation.\vspace{-0.3cm}

\begin{table}[!h]
\centering
\setlength{\tabcolsep}{6pt}
\caption{Data observation counting on works.}\label{tab:data}
\begin{tabular}{lcccc} \hline
 & longest & shortest & average & \#\_all\_words \\ \hline
Original sentences & 1019 & 5 & 27.4 & 156,565 \\
Compression & 40 & 2 & 10.4 & 89,238 \\ \hline
\end{tabular}\vspace{-0.5cm}
\end{table}

We can observe that the longest sentence is more than 1000 words while the longest of compression is 40. Generally, compression compresses a lot of unnecessary words. The shortest sequences of both original and compression sentences are quite short. The trend of the number of all words is quite similar to the average, in which original sentences are nearly twice longer than compressions, showing that a good model can reduce to nearly half of the words.\vspace{-0.3cm}

\subsubsection{Data Alignment}
The original data includes pairs of sentences. To train our compression model, we followed Filippova et al. to align original sentences and compressed sentences \cite{Filippova_EMNLP_15}. This is because we formulate the compression task as a sequence labeling task, in which unnecessary words are marked by a label, i.e. deletion or no-deletion and our model learns to remove deletion words.\vspace{-0.5cm}
\begin{figure}[!h]
    \centering
	\includegraphics[width=1\textwidth]{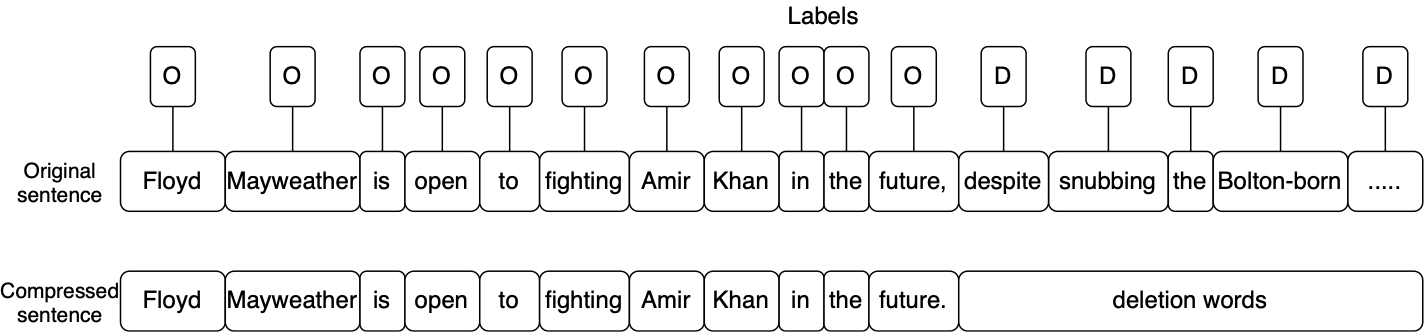}
	\caption{An alignment example. D denotes deletion and O is for no-deletion words.}\label{fig:alignment}\vspace{-0.5cm}
\end{figure}

The alignment was done in two steps. The first step is word segmentation, which segments a sentence into a set of words. The second step aligns words of original and compressed sentences. If a word in the original sentence appears in the gold compressed sentence, this word is kept (no-deletion); otherwise, it is deleted (deletion). Fig. \ref{fig:alignment} shows an example of the alignment.

We can observe that miss-matched words are labeled by \texttt{"D"}, showing that these words should be deleted from the compression model. After alignment, original sentences with labels are input for our model.\vspace{-0.2cm}

\subsection{The Proposed Model}
As mentioned, we formulated the compression as a sequence labeling task. Given an input sequence of words, our model learns to predict whether a word should be deleted or not. More precisely, this is a binary classification task. Given a word, our model needs to classify this word into deletion or non-deletion labels.\vspace{-0.3cm}

\begin{figure}[!h]
    \centering
	\includegraphics[width=1\textwidth]{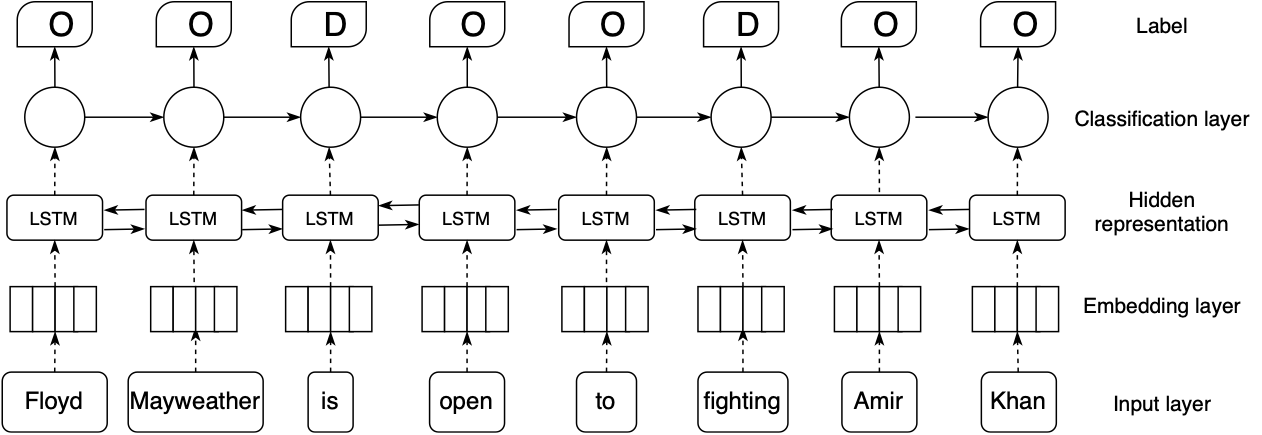}\vspace{-0.1cm}
	\caption{The overview of our model. The model outputs a label for each word. We put two incorrect labels (D) to show that in training, these output labels were used to compute the loss with the ground-truth data in Fig. \ref{fig:alignment}.}\label{fig:model}\vspace{-0.4cm}
\end{figure}

From this formulation, we introduce our model in Fig. \ref{fig:model}. The main idea of our model is that it exploits the efficiency of pre-trained language models for the embedding layer, and then stacks with BiLSTM for learning hidden representation from input sentences. The final layer is CRFs for classification to output a label for each word. We share the idea of stacking BiLSTM and CRFs with previous work because this architecture has achieved promising results in the task of named entity recognition \cite{Lample_NAACL_16,Ju_NAACL_18}. However, our model distinguishes to previous work in two points. Firstly, we exploit the power of pre-trained contextual transformers (Flair or BERT) for mapping tokens into low-dimensional vectors, instead of using word embeddings (Word2Vec or Glove) \cite{Lample_NAACL_16,Ju_NAACL_18}. It enables our model to encode the context aspect of a sequence into the learning process. Secondly, we design a flat structure, instead of using a nested structure \cite{Ju_NAACL_18}. It facilitates the learning process in a simple fashion. Our model also shares the idea of deleting unnecessary words as Filippova et al. \cite{Filippova_EMNLP_15}; however, exploiting pre-trained contextual embeddings from transformers makes a different point of our model compared to \cite{Filippova_EMNLP_15}. Also, we add one CRF layer for classification, instead of directly using softmax. It allows our model utilizing global optimization on a whole sequence. We next describe each layer of our model.\vspace{-0.2cm}

\subsubsection{Embedding layer}
The embedding layer is to use for mapping tokens into low-dimensional vectors. It is possible to employ any embedding method such as Word2Vec, Glove, or fastText; however, we utilized pre-trained contextual embeddings from transformers. This is because such transformers were trained with the consideration of the context of tokens on a huge amount of data. As a result, these models output high-quality embeddings compared to other word embedding methods (Glove). Inspired by the recent success of transformers in many NLP tasks \cite{Akbik_COLING_18,Devlin_NAACL_19}, we exploited Flair and BERT as our embedding layer.\vspace{-0.2cm}

\paragraph{\textbf{Flair}}
Flair is contextual string embeddings, which were from a character language model to produce word embeddings \cite{Akbik_COLING_18}. Flair has two distinct properties. Firstly, embeddings were trained without any explicit notion of words, hence, words were basically modeled in the form of sequences of characters. Secondly, embeddings were contextualized by their surrounding texts, meaning that the same word has different meanings (different word vectors) depending on its context. We take into account the contextual aspect from Flair to our model by using embeddings from Flair to map words in low-dimensional vectors for learning.\vspace{-0.2cm}

\paragraph{\textbf{BERT}}
(Bidirectional Encoder Representation from Transformer) is a multi-layered bidirectional Transformer encoder, which allows our model to represent the context of a word by considering its neighbors \cite{Devlin_NAACL_19}. The key idea of BERT is that it applies the bidirectional training of Transformer combined with multi-head attention to learn contextual relationships among words in a sequence. It includes two separate mechanisms: an encoder that reads the text input and a decoder that produces a prediction given an input sequence. Due to the purpose of building a language model, BERT only needs the encoder.

Different from directional models, which read the text input sequentially (left-to-right or right-to-left), the transformer encoder of BERT reads the entire sequence of words. With the combination of a large number of layers and multi-head attention, BERT can deeply capture the meaning of words based on their context. For example, considering the word ``bank” in two sentences ``I went to the bank to deposit some money” and ``I went to the bank of the river”, the representations of ``bank” are identical for a left-context unidirectional model, but they are distinguished with the use of BERT. This characteristic is appropriate with our task, in which our model needs to understand the context of a sequence for the deletion of words, i.e. detecting deletion or no-deletion words.

We employed a pre-trained BERT as the embedding layer due to two reasons: (i) BERT has achieved promising results on 11 NLP tasks, including sequence tagging \cite{Devlin_NAACL_19,Nguyen_PACLING_19} and (ii) BERT was contextually trained on a huge amount of data, so it can encode word meaning, which is important for machine learning models.\vspace{-0.2cm}

\subsubsection{Hidden Representation}
The hidden representation layer receipts output from the embedding layer to learn the hidden representation of input tokens. To do that, we stacked a layer by using LSTM \cite{Hochreiter_Schmidhuber_97} on the embedding layer.

LSTM is a variation of recurrent neural networks. It was designed to tackle long-term dependency of words and vanishing gradient descent in training deep learning models. Each LSTM cell uses the gate mechanism to decide whether input information is kept or ignored. The gate mechanism allows our model capturing the hidden representation of a long sequence. In practice, we employed  BiLSTM for training our model. This is because BiLSTM encodes information from forward and backward directions, showing higher efficiency in learning data representation compared to LSTM \cite{Ju_NAACL_18,Devlin_NAACL_19}.\vspace{-0.3cm}

\subsubsection{Classification}
The final layer is classification to predict labels of each word represented from the hidden presentation layer. In order to do that, we utilized CRFs as the final layer for prediction. This is because CRFs were designed for the sequence labeling task \cite{Lafferty_Pereira_ICML_01} and stacking CRFs on the top of BiLSTM has achieved promising results for the task of named entity recognition \cite{Lample_NAACL_16,Ju_NAACL_18}.

CRFs were designed to globally predict label sequences of any given sequences. Given an input sequence $X = (x_1, x_2, ..., x_n)$ which is the output from the BiLSTM layer, CRFs learn to make the prediction of $x_i$ by maximizing the log probability during training. More precisely, given a sentence sequence $X = (x_1, x_2, ..., x_n)$ and the corresponding state of sequence (labels) $Y = (y_1, y_2, ..., y_n)$, the probability of $Y$ conditioned on $X$ was defined as $P(Y|X)$. When making prediction of $y_i$, CRFs consider the current input $x_i$ (the current word) and previous states of previous words as $P(y_i = 1 |X)$. For example, in Fig. \ref{fig:model}, the label of \texttt{``fighting"} is decided by using information from previous words. In practice, we consider the last one or two steps because using all information has a higher computational cost for training and inference.\vspace{-0.3cm}

\subsubsection{Training}
Our model was trained in an end-to-end fashion. The classification layer predicts a label of each word. Predicted labels were used to compare to ground-truth labels of training data to compute the cross-entropy loss. Error loss was updated for training by using back-propagation.

We used the BERT-large cased model trained for English with the BookCorpus (800M words) and English Wikipedia (2,500M words) as mentioned in \cite{Devlin_NAACL_19}. For Flair, we used the English language models in both forward and backward directions trained on 1-billion words corpus mentioned in \cite{Akbik_COLING_18}. To increase the representation, we concatenated word vectors from Glove to word vectors from Flair or BERT in the embedding layer. This is because we want to increase the data representation of our model for learning.\vspace{-0.2cm}

\section{Settings and Evaluation Metrics}\label{sec:setting}\vspace{-0.1cm}
\subsection{Settings}
We used a new version of the dataset, which includes 200,000 pairs for training our model. We randomly selected 1000 pairs \cite{Filippova_EMNLP_15,Wang_ACL_17,Zhao_ACL_18} from 10,000 pairs of human annotation for automatic evaluation and the rest for validation. We used Adam optimizer for training in 100 epochs with gradient clipping at 5, except for BERT-MLP in which we found performed best with 10 epochs of training. LSTM layers have 256 hidden states. For Flair's language models, each composed of one-layered LSTM with 2048 hidden states. The BERT-large model has 24 layers, a hidden layer of 1024 neurons, 16 heads and 340M parameters.\vspace{-0.2cm}

\subsection{Evaluation Metrics}
We used F-score and compression rate for evaluation \cite{Filippova_EMNLP_13,Filippova_EMNLP_15} as follows.\vspace{-0.2cm}

\subsubsection{F-score}
We used recall and precision to compute the F-1 of compression and ground-truth data. These metrics were computed as follows.
\begin{align*}
    R = \frac{\#correct\_del\_words}{\#all\_del\_words}; \quad
    P = \frac{\#correct\_del\_words}{\#del\_words}; \quad
    \text{F-1} = \frac{2 \times R \times P}{R+P}
\end{align*}{}\vspace{-0.2cm}
    


where: $\#correct\_del\_words$ is the number of correct deletion words from the compression model compared to the ground-truth data; $\#all\_del\_words$ is all deletion words in ground-truth data; and $\#del\_words$ is the number of deletion words processed by the compression model.\vspace{-0.3cm}

\subsubsection{Compression Rate}
considers the length aspect of compression. It was computed by dividing the length of the compression over the original sentence length.\vspace{-0.3cm}

\subsection{Baselines}\vspace{-0.1cm}
We compared our model to several strong methods of sentence compression. F-score of several models can be found in the leader board (\texttt{http://nlpprogress.com/\\english/summarization.html}).
\begin{itemize}
    \item \textbf{ILP+features:} uses Integer Linear Programming with a set of features to find out optimal paths on syntactic trees \cite{Filippova_EMNLP_13}. The compressions are sub-trees generated from the optimization algorithm.
    
    \item \textbf{BiRNN+LM:} uses a language model as an evaluator to supporting BiRNN trained by a reinforcement learning fashion for sentence compression \cite{Zhao_ACL_18}.
    
    \item \textbf{LSTM:} is the extension work of \cite{Filippova_EMNLP_13}. The authors treated compression a deletion task \cite{Filippova_EMNLP_15}. To do that, the authors used LSTM to deal with the sequence labeling task, in which the model tries to delete unnecessary words.
    
    \item \textbf{BiLSTM:} is the same with LSTM but using a bi-directional LSTM cell for learning hidden representation \cite{Wang_ACL_17}.
    
    \item \textbf{BiLSTM-CRF:} We implemented BiLSTM-CRF as a baseline of our model. It uses Glove for the embedding layer and BiLSTM to learn hidden representation. The final layer uses CRFs for classification. This model does not use Flair or BERT embeddings. 
    
    \item \textbf{BERT+MLP:} We utilized BERT as the embedding layer and multi-layer perceptron (MLP) for classification. 
    
    \item \textbf{BERT+BiLSTM-CRF:} It uses BERT as an embedding layer and stacks BiLSTM. It finally uses CRFs for classification to detect unnecessary words.\vspace{-0.3cm}
\end{itemize}{}

\section{Results and Discussion}\label{sec:result}\vspace{-0.1cm}
This section first reports the comparison of our model with baselines. It next shows the observation of embeddings and error analysis.\vspace{-0.3cm}

\subsection{F-score Comparison}
\vspace{-0.1cm}
Table \ref{tab:result} shows the comparison of our model and baselines. We can observe that our model is better than baselines in terms of F-score with quite large margins. For example, our model is significantly better than BiLSTM-CRF (0.889 vs. 0.820 with $p$-value $\leq0.05$ with pairwise $t$-test)\footnote{https://docs.scipy.org/doc/scipy-0.19.0/reference/generated/scipy.stats.ttest\_ind.html.} even they share the same hidden representation and classification layers. The improvement comes from two points. Firstly, we exploit contextual embeddings from transformers instead of directly using word embeddings (Glove) for the embedding layer. Using these embeddings allows our model taking into account the contextual aspect of words trained on a huge amount of data. From that, our model can correctly distinguish necessary and unnecessary words of input sequences. This again confirms the efficiency of transformers for NLP tasks \cite{Akbik_COLING_18,Devlin_NAACL_19}. By contrast, BiLSTM-CRF directly uses Glove as the embedding layer, which somehow limits the model in learning context among words. Another reason is that we also exploit the power of word embeddings by concatenating word vector from Glove with vector generated from transformers. This combination enables our model to increase data representation, leading to improvements.\vspace{-0.4cm}

\begin{table}[!h]
\centering
\setlength{\tabcolsep}{5pt}
\caption{Comparison of sentence compression models. $^{\dagger}$ shows that our model is significantly better with $p$-value $\leq0.05$.}\label{tab:result}
\begin{tabular}{lccc} \hline
Model & Embedding & F-score & Compression Rate \\ \hline
ILP+features \cite{Filippova_EMNLP_13} & --- & 0.843 & --- \\
BiRNN+LM \cite{Zhao_ACL_18} & --- & 0.851 & 0.39 \\
LSTM \cite{Filippova_EMNLP_15} & Word2Vec & $0.820^{\dagger}$ & 0.38 \\
BiLSTM \cite{Wang_ACL_17} & --- & $0.800^{\dagger}$ & 0.43 \\
BiLSTM-CRFs & Glove & $0.820^{\dagger}$ & 0.40 \\ 
BERT(original)+MLP & BERT & 0.867 & 0.38 \\ \hline
BERT+BiLSTM-CRF (Ours) & BERT+Glove & \textit{0.883} & 0.39 \\
Flair+BiLSTM-CRFs (Ours) & FLair+Glove & \textbf{0.889} & 0.39 \\ \hline
\end{tabular}\vspace{-0.4cm}
\end{table}

For variants of LSTM (BiRNN, LSTM, and BiLSTM), our model still achieves better F-score from 3-5 percentage points. This supports our idea that using contextual embeddings for word representation. Compared to BiLSTM and BiLSTM-CRF, the model using CRFs for prediction is 2 percentage points better than that of only using BiLSTM. This shows the contribution of CRFs for making global label sequence prediction. The ILP model obtains a promising F-score because it uses features for weighting concepts. The features help to increase the quality of deletion based on constraints. Compared to the original BERT, our model is still better, but the margins among compression models are small. This is because the original BERT takes into account the contextual aspect of words for learning. We can also observe the contribution of BiLSTM-CRF, in which BERT with this architecture gives a better F-score than BERT using MLP (Tables \ref{tab:result} and \ref{tab:glove}). The model using Flair is slightly better than that using BERT (0.889 vs. 0.883). A possible reason is that BERT may be appropriate for quite long documents while the dataset includes short sentences. As a result, the BERT-based method is challenged to encode short separated texts. However, the gap is tiny, showing that the contextual embeddings from BERT can still contribute to our model.

The compression rate (CR) of compression models is quite similar, even they yield different F-scores. BiLSTM outputs the longest sequences, followed by BiLST-CRFs. Our model is not the best in terms of CR, compared to BERT+MLP. A possible reason may come from BiLSTM because BiLSTM-based methods tend to output longer sequences compared to LSTM or RNN.

To avoid over-fitting, we tested our model five times on the test set. In each time, we randomly selected 1000 samples from 10,000 pairs created by humans for testing and the rest for validation. We also used the same data segmentation to train BiLSTM-CRFs and BERT+MLP. We did not compare to other methods due to the different setting. The F-score was the average of our model in five times.\vspace{-0.4cm}
\begin{table}[!h]
\centering
\setlength{\tabcolsep}{5pt}
\caption{The F-score of testing three models in five times.}\label{tab:result-5times}
\begin{tabular}{lccc} \hline
Model & Embedding & F-score & Compression Rate \\ \hline
BiLSTM-CRFs & Glove & $0.820^{\dagger}$ & 0.40 \\ 
BERT(original)+MLP & BERT & 0.863 & 0.38 \\
Flair+BiLSTM-CRFs (Ours) & FLair+Glove & \textbf{0.887} & 0.39 \\ \hline
\end{tabular}\vspace{-0.4cm}
\end{table}
Table \ref{tab:result-5times} shows that our model consistently achieves better F-scores than BiLSTM-CRFs and the original BERT after testing in five times. This again confirms the efficacy of our model. Comparing to results in Table \ref{tab:result}, F-scores are slightly different due to the average on five times. \vspace{-0.3cm}

\subsection{Glove Contribution}\vspace{-0.1cm}
As mentioned, we concatenated embeddings from Glove to word vectors from Flair and BERT. To observe the contribution of Glove, we compared contextual-based compression models using with/without word vectors from Glove. \vspace{-0.4cm}

\begin{table}[!h]
\centering
\setlength{\tabcolsep}{5pt}
\caption{The contribution of Glove.}\label{tab:glove}
\begin{tabular}{lccc} \hline
Model & Embeddings & F-score & Compression Rate \\ \hline
BERT(w/o Glove)+BiLSTM-CRF & BERT+Glove  & 0.875 & 0.39 \\
BERT(with Glove)+BiLSTM-CRF & BERT+Glove & \textbf{0.883} & 0.39 \\ \hline
Flair(w/o Glove)+BiLSTM-CRF & Flair+Glove & 0.877 & 0.38 \\
Flair(with Glove)+BiLSTM-CRF & Flair+Glove & \textbf{0.889} & 0.39 \\ \hline
\end{tabular}\vspace{-0.4cm}
\end{table}

From Table \ref{tab:glove}, we can observe that models using Glove obtain better F-cores than those which do not use Glove. This is because the concatenation of embeddings from Glove and contextual embeddings enriches the data representation of sentences, hence, it improves the learning quality of our model. However, the margins are small, showing the efficiency of contextual embeddings from Flair and BERT \cite{Akbik_COLING_18,Devlin_NAACL_19}. The compression rate of transformer variants is similar, which is consistent with the compression rate of transformer-based models in Table \ref{tab:result}.\vspace{-0.4cm}


\subsection{Output Observation}\vspace{-0.1cm}
We observed the outputs of (i) BiLSTM-CRF, (ii) BERT+MLP, (iii) BERT+BiLSTM-CRF, and (iv) our model. We can observe that for the first output (the half of the table), all models give correct compression. A possible reason is that the input sequence is quite short and simple, so all the models can correctly predict deletion words.\vspace{-0.35cm}
\begin{table}[!h]
\centering
\setlength{\tabcolsep}{2pt}
\caption{Output samples. Incorrect phrases are mark by strike-through words.}\label{tab:output}
\begin{tabular}{c p{10.3cm}} \hline
Original & Vine, the mobile app owned by Twitter, has banned sexually explicit content, effective immediately. \\ 
 Gold-comp & Vine has banned sexually explicit content. \\ \hline
(i) & Vine has banned sexually explicit content. \\
(ii) & Vine has banned sexually explicit content. \\
(iii) & Vine has banned sexually explicit content. \\
Our model & Vine has banned sexually explicit content. \\ \hline \hline

Original & A man found dead in a Fairfield hotel room on Sunday, Sept. 1 has been identified as Matthew Wuss, 20, of Chester. \\ 
 Gold-comp & A man found dead in a Fairfield hotel room has been identified. \\ \hline
(i) & A man found dead in a Fairfield hotel room \sout{on Sept. 1} has been identified \sout{as Matthew Wuss}. \\
(ii) & A man found dead in a Fairfield hotel room has been identified \sout{as Matthew Wuss}. \\
(iii) & A man found dead in a Fairfield hotel room \sout{on Sunday} has been identified. \\
Our model & A man found dead in a Fairfield hotel room has been identified. \\ \hline
\end{tabular}\vspace{-0.4cm}
\end{table}
For the second output, our model outputs correct compression. This supports our idea in exploiting contextual embeddings for word representation and confirms F-scores in Table \ref{tab:result}.  BiLSTM-CRF shares one incorrect phrase (\texttt{"as Matthew Wuss"}) with BERT-MLP. The reason may come from the use of BiLSTM-CRF, which may be limited for learning data representation of long sequences. In addition, BiLST-CRF does not use contextual embeddings, so it makes a challenge to capture the meaning of words from input sequences. As a result, it predicts two wrong phrases. It again confirms F-scores in Table \ref{tab:result}, in which the BiLSTM-CRF compression model yields a quite low F-score.\vspace{-0.2cm}

\section{Conclusion}\label{sec:conclusion}\vspace{-0.1cm}
This paper introduces a model for sentence compression as deletion. Our model utilizes the power of contextual embeddings for capturing the context aspect of input words for learning. The embeddings combined with BiLSTM and CRF allow our model efficiently learning hidden representation from data. Experimental results on a benchmark dataset show two points. Firstly, contextual embeddings contribute to the compression model. This leads to a new state-of-the-art F-score on the Google dataset. Secondly, by adding word vectors from Glove, our model improves the quality of deletion compared to the model without using Glove.

One possible direction is to analyze syntactic features that can be integrated into our model. Also, parsed tree representation may be helpful for deep learning.\vspace{-0.25cm}

\section*{Acknowledgement}\vspace{-0.1cm}
This research is funded by Hung Yen University of Technology and Education under the grant number UTEHY.L.2020.04.\vspace{-0.2cm}

%
%
%

\begin{thebibliography}{10}
\providecommand{\url}[1]{\texttt{#1}}
\providecommand{\urlprefix}{URL }
\providecommand{\doi}[1]{https://doi.org/#1}

\bibitem{Akbik_COLING_18}
Akbik, A., Blythe, D., Vollgraf, R.: Contextual string embed-dings for sequence labeling. In: Proceedings of the 27th International Conference on Computational Linguistics, pp. 1638-1649 (2018)

\bibitem{Kirkpatrick_ACL_11}
Berg-Kirkpatrick, T., Gillick, D., Klein, D.: Jointly learning to extract and
  compress. In: Proceedings of the 49th Annual Meeting of the Association for
  Computational Linguistics: Human Language Technologies-Volume 1, pp. 481-490 (2011)

\bibitem{Cho_EMNLP_14}
Cho, K., van Merrie~̈nboer, B., Gulcehre, C., Bahdanau, D., Bougares, F.,
  Schwenk, H., Bengio, Y.: Learning phrase representations using rnn
  encoder-decoder for statistical machine translation. In: Proceedings of the
  2014 Conference on Empirical Methods in Natural Language Processing, pp.
  1724-1734 (2014)

\bibitem{Clarke_JAIR_08}
Clarke, J., Lapata, M.: Global inference for sentence compression: An integer
  linear programming approach. J. Artif. Intell. Res. \textbf{31}, 399-429  (2008)

\bibitem{Devlin_NAACL_19}
Devlin, J., Chang, M.W., Lee, K., Toutanova, K.: Bert: Pre-training of deep
  bidirectional transformers for language understanding. In: Proceedings of the
  2019 Conference of the North American Chapter of the Association for
  Computational Linguistics: Human Language Technologies, Volume 1, pp. 4171-4186 (2019)

\bibitem{Filippova_EMNLP_15}
Filippova, K., Alfonseca, E., Colmenares, C.A., Łukasz Kaiser, Vinyals, O.:
  Sentence compression by deletion with lstms. In: Proceedings of the 2015
  Conference on Empirical Methods in Natural Language Processing, pp. 360-368
  (2015)

\bibitem{Filippova_EMNLP_13}
Filippova, K., Altun, Y.: Overcoming the lack of parallel data in sentence
  compression. In: Proceedings of the 2013 Conference on Empirical Methods in
  Natural Language Processing, pp. 1481-1491 (2013)

\bibitem{Hochreiter_Schmidhuber_97}
Hochreiter, S., Schmidhuber, J.: Long short-term memory. Neural computation \textbf{9}(8) , 1735-1780  (1997)

\bibitem{Ju_NAACL_18}
Ju, M., Miwa, M., Ananiadou, S.: A neural layered model for nested named entity
  recognition. In: Proceedings of the 2018 Conference of the North American
  Chapter of the Association for Computational Linguistics: Human Language
  Technologies, Volume 1 (Long Papers), pp. 1446-1459 (2018)

\bibitem{Lafferty_Pereira_ICML_01}
Lafferty, J., McCallum, A., Pereira, F.: Conditional random fields:
  Probabilistic models for segmenting and labeling sequence data. In:
  Proceedings of the 18th International Conference on Machine Learning, pp.
  282–289 (2001)

\bibitem{Lample_NAACL_16}
Lample, G., Ballesteros, M., Subramanian, S., Kawakami, K., Dyer, C.: Neural
  architectures for named entity recognition. In: NAACL-HLT, pp. 260-270 (2016)

\bibitem{Liu_SIGKDD_19}
Liu, C., Wang, P., Xu, J., Li, Z., Ye, J.: Automatic dialogue summary
  generation for customer service. In: Proceedings of the 25th ACM SIGKDD
  International Conference on Knowledge Discovery \& Data Mining, pp. 1957-1965
  (2019)

\bibitem{Nguyen_PACLING_19}
Nguyen, M.T., Phan, V.A., Linh, L.T., Son, N.H., Dung, L.T., Hirano, M., Hotta,
  H.: Transfer learning for information extraction with limited data. In:
  Proceedings of 16th International Conference of the Pacific Association for
  Computational Linguistics (2019)

\bibitem{Vanetik_IS_20}
Vanetik, N., Litvak, M., Churkin, E., Last, M.: An unsupervised constrained
  optimization approach to compressive summarization. Information Sciences \textbf{509},
  22-35  (2020)

\bibitem{Wang_ACL_17}
Wang, L., Jiang, J., Chieu, H.L., Ong, C.H., Song, D., Liao, L.: Can syntax
  help? improving an lstm-based sentence compression model for new domains. In:
  Proceedings of the 55th Annual Meeting of the Association for Computational
  Linguistics (Volume 1: Long Papers), pp. 1385-1393 (2017)

\bibitem{Zhao_ACL_18}
Zhao, Y., Luo, Z., Aizawa, A.: A language model based evaluator for sentence
  compression. In: ALC (Volume 2: Short Papers), pp. 170-175 (2018)

\end{thebibliography}
%

\end{document}